\documentclass{styles/svproc}
\usepackage{url}
\usepackage{graphicx}
\usepackage{wrapfig}
\usepackage[skip=2pt,font=footnotesize]{caption}
\usepackage{subcaption}
\usepackage[rightcaption]{sidecap}
\usepackage{url}
\usepackage{array}
\sidecaptionvpos{figure}{h}
\usepackage{listings}
\usepackage[colorinlistoftodos]{todonotes}
\begin{document}
\mainmatter              % start of a contribution
\title{ElasticBroker: Combining HPC with Cloud to Provide Realtime Insights into Simulations}
\titlerunning{ElasticBroker: Combining HPC with Cloud to Provide Realtime Insights}  % abbreviated title (for running head)
\author{Feng Li\inst{1} \and Dali Wang \inst {2}\and Feng Yan \inst{3} \and Fengguang Song \inst{1}
}
\authorrunning{F. Li, D. Wang, F. Song} % abbreviated author list (for running head)
\tocauthor{Feng Li, Dali Wang, Fengguang Song}
\institute{Indiana University Purdue University, Indianapolis, IN 46202, USA. \\
\email{li2251@purdue.edu}, \, \, \email{fgsong@cs.iupui.edu}
%\\ WWW home page:
%\texttt{http://users/\homedir iekeland/web/welcome.html}
\and
Oak Ridge National Laboratory, Oak Ridge, TN 37831, USA. \\
\email{wangd@ornl.gov}
\and
University of Nevada, Reno, NV 89557, USA \\
\email{fyan@unr.edu}
}

\maketitle              % typeset the title of the contribution

\begin{abstract}
% 250 words for SMC 2020
%Large-scale scientific simulations can generate significant amounts of data over time, and 
For large-scale scientific simulations, it is expensive to store raw simulation results to perform post-analysis.
To minimize expensive I/O, ``in-situ'' %or ``in-transit''
analysis is often used, where analysis applications are tightly coupled with scientific simulations and can access and process the simulation results in memory.
%while in-situ analysis shares the same computing resources with the primary simulation tasks, in-transit analysis typically uses dedicated resources for analysis.
Increasingly, scientific domains employ Big Data approaches to analyze simulations for scientific discoveries.
However, %HPC simulations and Big Data analysis frequently use significantly different data representations.
it remains a challenge to organize, transform, and transport data at scale between 
the two semantically different ecosystems
(HPC and Cloud systems).
In an effort to address these challenges, we design and implement the ElasticBroker software framework, which bridges HPC and Cloud applications to form an ``in-situ'' scientific workflow.
%parallel MPI-based applications (in HPC systems) and Big Data applications (in Cloud platforms) efficiently.
Instead of writing simulation results to parallel file systems, ElasticBroker performs data filtering, aggregation,
and format conversions to close the gap between an HPC ecosystem and a distinct Cloud ecosystem.
To achieve this goal, ElasticBroker reorganizes simulation snapshots into continuous data streams and send them to the Cloud.
In the Cloud, we deploy a distributed stream processing service
to perform online data analysis.
In our experiments, we use ElasticBroker  to setup and execute a cross-ecosystem scientific workflow, which consists of a parallel computational fluid dynamics (CFD) simulation running on a supercomputer, 
and a parallel dynamic mode decomposition (DMD) analysis application running in a Cloud computing platform.
Our results show that running scientific workflows consisting of decoupled HPC and Big Data jobs in their native environments with ElasticBroker,
can achieve high quality of service, good scalability, and provide high-quality analytics for ongoing simulations.  

%Those data are often stored in parallel file-systems first,
%copied to another site and then read again by different analytics applications for further investigation/exploration.
% We would like to encourage you to list your keywords within
% the abstract section using the \keywords{...} command.
\end{abstract}
\section{Introduction}

HPC and Big Data ecosystems are significantly different from each other, and are designed and manufactured for their own purposes, respectively.
In the HPC world, systems are designed for faster execution of large-scale parallel programs.
Hundreds of thousands of processes run across a large number of compute nodes.
Those compute nodes are high-end servers equipped with many CPU cores and large-size memories,
and are tightly connected by fast interconnects such as InfiniBand.
Simple and minimal operating system kernels and software stacks  are used in those computer nodes for efficient operation.
Often, the low-level, highly portable and efficient message-passing parallel programming model (MPI) is used,
such that processes in different address spaces can work collaboratively and talk with each other through point-to-point or collective communications.

%\todo{As a example, the S3D simulation, can run across xx nodes in xxx days. }

Big Data is a totally different world, where applications are designed to collect, process, and analyze large amounts of data to gain knowledge.
Software in a Big Data ecosystem such as Apache Spark or Hadoop can use the high-level MapReduce programming model
to execute data analysis jobs on clusters of commodity machines \cite{dean2008mapreduce}.
More recently, cloud computing technologies such as container and service-oriented architecture have further hidden the complexity of parallel software packages,
and have made Big Data platforms more accessible to developers.
Overall, the general architecture and design commonly found in Big Data ecosystems, help users process and analyze data at large scale affordably and reliably. 
More details of the comparison between the HPC and Big Data ecosystems have been discussed and presented by Reed and Dongarra \cite{reed2015scientific}. 
%Paper in PDF: https://dl.acm.org/doi/pdf/10.1145/2699414?download=true 

%removed, too simple to show it.
%Table \ref{tab:compare} summaries some of the fundamental differences between the HPC and Big Data ecosystems.

%\todo{Add a table to show the major differences of two ecosystems}
%\begin{table}
% \caption{Comparison of typical HPC systems and Big Data systems.\label{tab:compare}}
%\begin{center}
% \begin{tabular}{ | m{6em} | m{5cm}| m{5cm} | } 
% \hline
% Ecosystem & HPC & Big Data \\ [0.5ex] 
% \hline \hline
% Programming model & Message-passing & MapReduce  \\ 
%  \hline
% Software stack & simplified & complex \\
%  \hline
% Hardware & high-end nodes with fast interconnect & cluster of accommodate machines \\
%  \hline
%%Principles & FLOPS(speed) & accesibility, fault-tor \\ [1ex] 
% \hline
%\end{tabular}
%\end{center}
%\end{table}

In practice, many scientific computing applications not only have long execution time, but also generate ``big'' amounts of data.
For instance, peta-bytes of data may be generated from a single run of a scientific simulation.
The generated data is traditionally stored in a parallel file system, then copied to another site, and read again by different analysis applications for further investigation or exploration. Such a data storage/movement/post-analysis pattern can be extremely expensive, and hence there is an inevitable trend to pursue in-situ data analysis, where analysis applications can continuously process and analyze the in-memory data structures while the simulation applications are running  \cite{docan2012dataspaces,fu_performance_2018,dreher2017decaf,bennett2012combining}.
%are either tightly coupled with the simulation application on the same computer nodes, or continuously offloaded to different computing resources. 
%Our previous work Zipper runtime system[2] utilizes task parallelism and pipeline parallelism to couple applications, so that data generated by simulation processes can be consumed by analytics processes in nearly real-time.

However, existing in-situ analysis frameworks often run in the scope of HPC ecosystem.
%(in most cases, involving one HPC system).
The main problem is that almost all data analytics and machine learning (ML) applications have been written using Big Data programming languages and libraries (e.g., Python, Scala, MapReduce, Spark, etc.), 
and are often deployed to Cloud computing platforms.
In addition, these data analytics or ML software and tools have already been widely accepted by
the Big Data community, and fully tested on Cloud platforms/services such as Apache Spark or Google Dataflow.
%high-level language(e.g. python) and Big Data community have already provided great platform/services like Apache Spark/Google Dataflow.
Hence, the question is: can a system be designed that can run scientific workflows which consist of both native HPC and Big Data applications?
Since it is almost impossible to port all data analytics/ML libraries from the Big Data ecosystem to the HPC ecosystem (or vice versa), we strive to bridge the gap and integrate HPC with Cloud environments,
so that we may utilize the elastic services and native software in the Cloud to analyze HPC simulations efficiently.

There are several challenges to achieve the goal.
Firstly, the data formats between HPC systems and Cloud services are usually different.
It is a non-trivial task to offload data from HPC simulations to Cloud applications, and apply necessary data transformations correctly and efficiently.
Also, the bandwidth between HPC and Cloud systems is limited, and bridging services between the two ecosystems must be carefully designed to minimize the data transmission overhead.
Furthermore, the mapping between simulation processes and data analysis processes should be optimized to minimize data flow stalls.
To tackle those challenges, we present ElasticBroker, which bridges the ecosystems of HPC and Cloud.
When MPI-based HPC applications are linked with the ElasticBroker library,
%, instead of writing data into parallel file system for future analysis,
the simulation data will be transformed to Cloud-native data objects and continuously streamed to the data analysis services deployed in Cloud systems,
where the data objects together with scheme information are organized and analyzed.

To showcase our system, we develop a real-world cross-ecosystem scientific workflow, which has: 
\begin{itemize}
    \item a parallel MPI-based computational fluid dynamics (CFD) simulation running in HPC, and
    \item a distributed online Dynamic Mode Decomposition (DMD) application using stream processing service deployed in Cloud.
\end{itemize}

We build and execute this workflow on the IU Karst HPC \cite{iukarst}  and XSEDE Jetstream Cloud systems\cite{stewart_jetstream_2015,towns_xsede_2014}.
From the experimental results, we observe that by linking CFD applications with ElasticBroker, we can  effectively migrate the simulation data from HPC system,
and by using the remote Cloud analysis services, we can provide in-time insights into the ongoing fluid dynamics.
\section{Background}
In this section, we first introduce Cloud-based stream processing.
Then, we present the background knowledge of Dynamic Mode Decomposition, which is an analysis method we have deployed in our Cloud-based stream processing service. 

\subsection{Cloud-based stream processing data analytics}
Nowadays it has become common that data is generated continuously over time.
For example, sensor data generated from IoT devices or web logs are produced from multiple sources and can accumulate everyday.
Instead of storing the data and doing post-processing in future, stream processing can be used to give real-time insights of the data.
The advantage of being ``real-time" is essential in various scenarios such as online fraud detection and emergency handling, where it can help early decision-making.

In stream processing, ``unbounded'' datasets (or ``data streams'') are used as input.
New data records are continuously added to those data streams, where they can be analyzed on the fly.
Popular stream processing frameworks (e.g., Apache Kafka \cite{kreps2011kafka}, Flink \cite{carbone2015flink}, Storm\cite{apachestorm}, and Spark Streaming \cite{sparkstreaming})
have been extensively used in different areas to provide in-time analytics for various data sources. 
Popular Cloud providers now offer data analytics as a service (e.g., Google DataProc \cite{googledataproc}, Amazon Kinesis Streams \cite{amazonkinesis}),
so that users can interact with the service using their favorite programming languages %(e.g. Python),
regardless of platform infrastructure. 

In the case of computational fluid dynamics (CFD) in the HPC domain, the simulations can run over days or even months.
%Real-time analysis of simulation results can help scientists discover  of ongoing simulations.
Analysis of data generated while the simulation is in progress can help scientists discover patterns and understand behaviors, which they would otherwise have to wait till the simulation finishes.
% \todo{example of realtime detection?} 
In this work, we explicitly utilize the convenience and advantages of Cloud-based stream processing to provide timely insights to the running simulations.

\subsection{Dynamic Mode Decomposition}
In fluid dynamics, the flow fields are organized in a complex, high-dimensional dynamical system.
It is well known that important flow features can be recognized through visual inspections of the flow, even when there are perturbations and variations \cite{taira2017modal}.
This means that some coherent structures exist in the fluid fields, which contain useful dynamical information of the fluids and can help researchers understand the patterns/behaviors of the fluid flows.
To mathematically extract those coherent structures from such dynamical systems, 
modal analysis techniques,  such as Dynamic Mode Decomposition analysis (DMD \cite{schmid2010dmd}), are often used.
Specifically, DMD analysis relies solely on snapshots (or measurements) of a given system, and provides the spatial-temporal decomposition of those data into a set of dynamical modes \cite{kutz_dynamic_2016}.
Since DMD is data-driven and doesn't need to model the governing equations of the fluids, it is considered as an ``equation-free'' and ``data-driven'' method.
Traditionally, DMD analysis has been used to study fluid structures from dynamic flow geometries \cite{rowley2009spectral}.
%Overtime, it has also been used in other domains, such as analyzing the spatial-temporal dynamics of the infectious disease data and extracting useful information from both historical and experimental disease data \cite{proctor_discovering_2015}.
In this work, we use DMD as an analysis example, and show how it can be deployed in the Cloud as a part of the distributed stream processing service, to analyze CFD simulations at real time.

\section{Methodology}

In this section, we present the design of our software framework and the
decisions we have made to solve the challenges of offloading analytical tasks to Cloud systems from the running simulations. The ElasticBroker framework contains two major components:
\begin{enumerate}
    \item A C/C++ brokering library in HPC, which transforms data from a simulation-domain format to a Cloud-compatible format.
    \item A distributed stream processing analysis service deployed in Cloud.
\end{enumerate}
Between HPC and Cloud, data is converted from the simulation by ElasticBroker, and then transferred to the distributed stream processing analysis service using available inter-site bandwidth.

%In the following of the this section, we will unveil more details of how we design $ElasticBroker$ and show how we address the above challenges. 
    
\subsection{HPC components}
\label{sect:hpc-components}
On the HPC side, %in-memory data structure from simulations will be converted into Cloud formats of objects.
%For example, a MPI\_IO\_Write operation will be translated to a keyvalue put operation.
commonly used I/O libraries, such as MPI-IO \cite{thakur_data_1999} and ADIOS \cite{lofstead_flexible_2008}, provide interfaces to output simulation data to the file system.
We provide a similar interface for HPC applications so that it is easy for existing simulation code to adapt to the ElasticBroker services, as shown in Listing \ref{lst:broker}.

\begin{figure}
	\centering
	\includegraphics[scale=0.5]{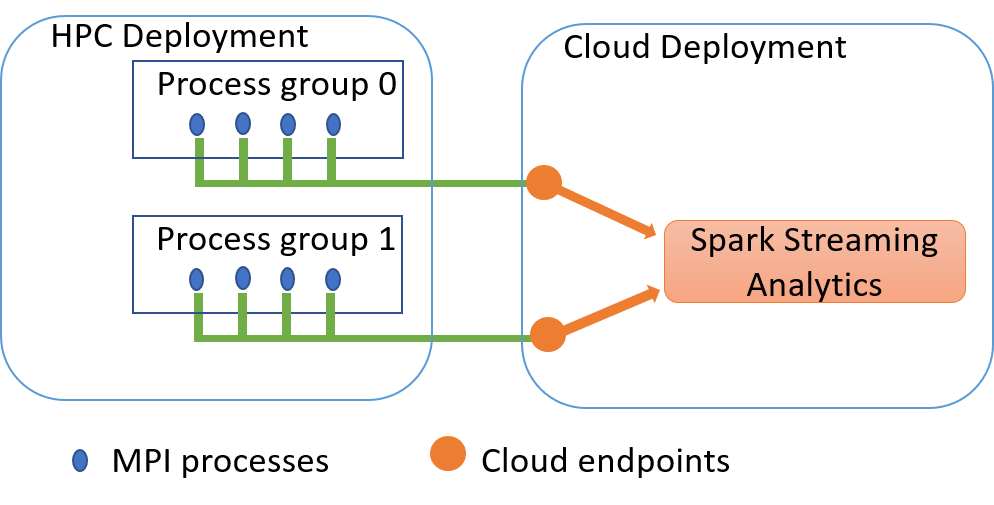}
	\caption{Process group in HPC and its relation with Cloud endpoints.
		In this example, MPI processes are divided into 2 groups (each with 4 processes),
		and process in one group will send its own data streams to one Cloud endpoint.}
	\label{fig:hpc-components}
\end{figure}

In Listing \ref{lst:broker}, a Cloud service consists of several endpoints.
Each of them is specified using $service\_ip$ and $service\_port$.
The {\tt broker\_init} function initializes the connections between HPC and Cloud by registering data fields from the simulation with remote Cloud service endpoints.
Those data fields are differentiated by the $field\_name$ variable in the API, such as ``pressure'' or ``velocity\_x''.
We divide the MPI processes in a simulation into groups (using $group\_id$), so that processes in one group will register themselves with one corresponding Cloud endpoint for future writes, as shown in Figure \ref{fig:hpc-components}.
During the main iterations of the simulation, the {\tt broker\_write} function is called iteratively, to transform field data from the simulation process into stream records, which are sent to the Cloud endpoint that process has connected to.
Each stream record contains the time-step information and the serialized field data of the simulation process.
In the Cloud, stream records received from all endpoints will be indexed, aggregated and partitioned by the stream processing service, which will be discussed in Section \ref{sect:cloud-components}.
%Then, data streams from one process group are directed to the streaming-processing engine deployed in Cloud.
%For simplicity, we let each MPI process(from Computer nodes) send data out directly through tcp socket. 
%Actually this can be improved by utilizing the HPC-side network topology(e.g. romio \cite{thakur_data_1999})

\begin{lstlisting}[language=C,caption=The ElasticBroker C/C++ API used by HPC applications., basicstyle=\small, label=lst:broker]
struct CloudEndpoint{
    char* service_ip;
    int service_port;
};

// A Cloud service can provide several endpoints
struct CloudEndpoint endpoints[NUM_GROUPS];

// Initialize the broker service, by connecting each MPI 
// process with one of the Cloud endpoints.
broker_ctx* broker_init(char* field_name, int group_id);

// write a chunk of in-memory data (void* data) to the broker.
broker_write(broker_ctx* context, int step, void* data,
size_t data_len);

// finalize the broker services
broker_finalize(broker_ctx* ontext);
\end{lstlisting}

Dividing HPC processes into groups enables us to assign each group to a designated Cloud endpoint for achieving a higher data transfer rate, as shown in Figure \ref{fig:hpc-components}.
%With large scale simulations, data generated from all MPI processes can be redirected to multiple endpoints.
Process groups also provide a higher degree of flexibility.
Users can decide how many endpoints are necessary based on the outbound bandwidth of each HPC node and inbound bandwidth of each Cloud endpoint.
%For now, we utilize a static mapping between HPC process groups and Cloud endpoints. In the future, we plan to add load balancing features to the cloud side.

\subsection{Cloud-based data analysis components}
\label{sect:cloud-components}
%When data are streamed to the Cloud through one of the endpoints, the streaming-processing engine deployed in Cloud will analyze the newly-added data.  
In this subsection, we will first introduce how we setup the Cloud stream processing service, and then describe how different components in the Cloud service work together to provide insights to the incoming streamed simulation data.

\subsubsection{Preparation}
We choose {\tt Spark Streaming} \cite{sparkstreaming} as our stream processing engine, which supports scalable, high-throughput, fault-tolerant stream processing of live data streams. 
By utilizing the core Spark functionality, we can apply basic operations such as map, reduce, filter, join, and advanced algorithms using Spark Machine Learning and Graph Processing libraries to data streams.
Currently, we deploy our Spark cluster and Cloud endpoints within a Kubernetes cluster in the Jetstream Cloud.
%We created this Kubernetes cluster using the Openstack Magnum Plugin.
As a popular container orchestration system, Kubernetes provides an abstraction layer above different Cloud providers \cite{bernstein2014containers}.
This way our stream processing setup can be easily reproduced with different Cloud providers like Google Cloud Platform or Amazon AWS.

%\todo{explain trigger time}

\begin{figure}
	\centering
	\includegraphics[scale=0.5]{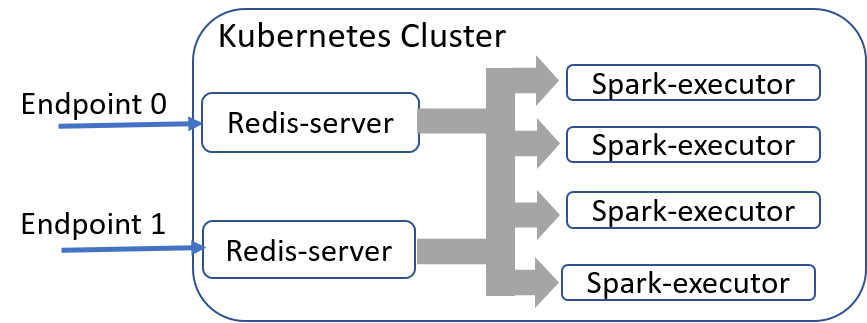}
	\caption{The deployment of our Cloud components. Each Redis-server container acts as an endpoint, and exposes the same TCP port to outside. %to which MPI process can stream simulation data. 
		The Spark-executor containers will read available data streams from Redis-server containers. All containers are scheduled in the Kubernetes cluster deployed in Jetstream, and use the in-cluster network to communicate with each other.}
	\label{fig:cloudcomponent}
\end{figure}

Figure \ref{fig:cloudcomponent} shows the overall layout of such Cloud settings.
Currently we use Redis server instances as our Cloud endpoints.
Redis, which is an in-memory data structure store, is used to accept data streams from the HPC components.
We use {\tt spark-redis} connector \cite{spark-redis} to let the Redis instances forward structured data to Spark stream processing services deployed in Cloud. 
All Redis instances export TCP port $6379$ to the outside of the Cloud system.
All of our Cloud services (Spark stream processing engine and Redis server instances) are containerized and are scheduled using Kubernetes's native scheduling, which makes it easy to adapt to different Cloud providers. 
Specifically, a Spark-executor container is comprised of 
the Python-based DMD analysis library PyDMD \cite{demo2018pydmd}, and related Scala software packages such as spark-redis. 
More details about the software we use in the Cloud services
are provided in Section \ref{sect:exp}.
We create and manage the Kubernetes cluster from a ``gateway'' VM (virtual machine) in Jetstream Cloud, using the Magnum Openstack Plugin \cite{magnum}.
After the cluster is set up, we use the {\tt spark-submit} command from the gateway VM, to launch the Spark-executor containers to the Kubernetes cluster.
%the packaged container path, and the destination Kubernetes cluster address, so that the Spark-executors will be scheduled to different Kubernetes nodes in Jetstream.

\subsubsection{Data Processing in Cloud}

\begin{figure}
	\centering
	\includegraphics[scale=0.5]{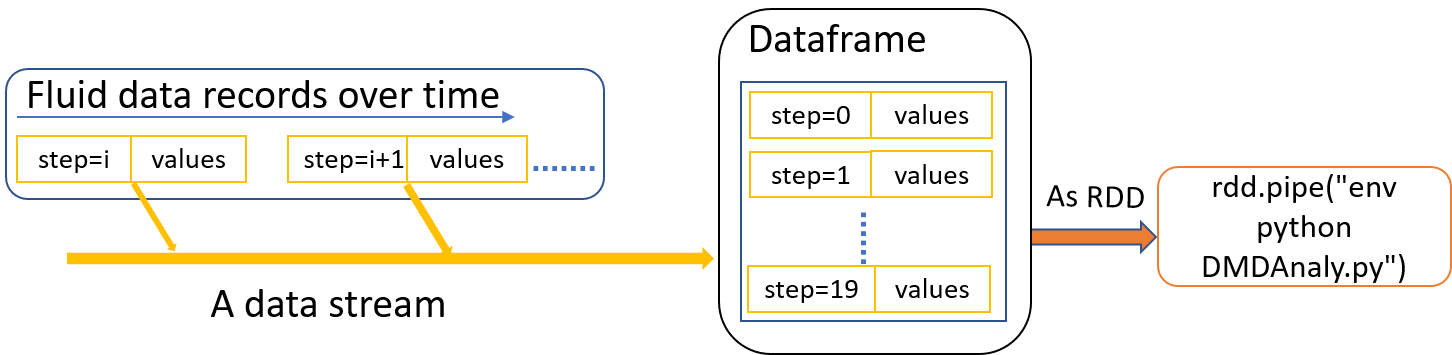}
	\caption{Data processing in the Cloud. Each MPI process sends data through a data stream, then unbounded data in each data stream is re-arranged into micro-batches (aka Spark Dataframes). Micro-batches from multiple data streams are treated as partitions of one Resilient Distributed Dataset (RDD). The {\tt rdd.pipe} function then sends each partition to the external Python script exactly once. }
	\label{fig:datasplit}
\end{figure}

When data is aggregated from different endpoints, Spark-executors will read records from data streams sent by all MPI processes.
%In the current settings, each MPI process initiates one data stream to one of the Cloud endpoints.
Fluid data (snapshots) from different simulation processes are added to the separate data streams over time.
Figure \ref{fig:datasplit} shows how data records in one data stream are aggregated as Spark ``Dataframes'', which are then processed by analysis code.

We let Spark manage the scheduling and parallelism, so that multiple executors can be mapped to different data streams and process the incoming data concurrently.
%Each data streams will be mapped to a partition of RDD.
%While the main programming environment in Spark is in Scala, the analytic code (PyDMD) is written in Python. 
%To direct data from Spark to external Python scripts,
We use the {\tt rdd.pipe} function \cite{sparkrddapi} in Spark to send Dataframe data from the main Spark context to external programs (in our case the Python interpreter).
This operation happens concurrently with all data streams, thanks to the design of Spark which enables a high degree of parallelism.
%When rdd.pipe is called, each spark-executor will open a Python session, and run the external PyDMD Python script using the formatted data from Dataframe as the standard input. 
The results of all Spark-executors are then collected using the {\tt rdd.collect} function so that they can be visualized/presented.
%Such practice can also be applied to other analytic services too.
%This way, we can utilize the capacity of Spark cluster to do data-intensive parallel work such as data filtering/grouping/sorting,
%while being able to export data into external programs written in various languages. 
\section{Experiments}
\label{sect:exp}
%We design and implement workflows that use Cloud stream processing to analyze simulation data.
We perform two sets of experiments to evaluate the 
performance of scientific workflows using ElasticBroker.
The first set of experiments use 
a real-world CFD simulation running in HPC, with DMD analysis deployed in Cloud, to show workflows with ElasticBroker can achieve good end-to-end time.
%We compare the simulation-side elapsed time and wor with the file-system based solution, to show that plugging with ElasticBroker has minimal performance effects on simulation application.
The second set of experiments use synthetic data generator processes in HPC and the same DMD analysis deployed in the Cloud
to evaluate ElasticBroker's throughput and qualify of service at different scales.

We use IU Karst as our HPC platform, which has specifications shown in Table \ref{tab:karstconfig}.
We deploy our Cloud analysis applications in XSEDE Jetstream Cloud \cite{stewart_jetstream_2015,towns_xsede_2014}, whose information is shown in Table \ref{tab:cloudconfig}. 
%Each of the Jetstream $m1.large$ instance we use has 10 vcpus and 30GB memory.
%Q: What is Jetstream's hardware information? Can you provide both hardware and software info for both systems?

\begin{table}[h]
    \begin{subtable}[h]{0.45\textwidth}
        \centering
        \begin{tabular}{l | l | l}
		 \hline
		 & Information \\
		 \hline
		CPU &
		 2 Intel Xeon \\
		 & E5-2650 v2.60 GHz CPUs\\
		\#cores &  16 cores per node\\
		Main memory & 32 GB per node\\
		Secondary storage &  250 GB per node\\
		Network & 10-gigabit Ethernet \\
		MPI & OpenMPI 2.1 \\
		Compiler version & GCC 5.4 \\
		OpenFOAM & v1906 \\
		 \hline
	\end{tabular}
       \caption{Karst System}
       \label{tab:karstconfig}
    \end{subtable}
    \hfill
    \begin{subtable}[h]{0.45\textwidth}
        \centering
        \begin{tabular}{l | l | l}
		 \hline
		 & Information \\
		 \hline
		VM image & Fedora-Atomic-29 \\ 
		VM type & m1.large \\
		VM vCPUs & 10 \\
		VM memory & 30GB \\
		Kubernetes version & 1.15.7\\
		Spark version & 2.4.5 \\
		Redis version & 5.0 \\
		 \hline
	    \end{tabular}
	    \caption{Jetstream System}
       \label{tab:cloudconfig}
       \end{subtable}
       \caption{Hardware and software information of the IU Karst HPC system and the XSEDE Jetstream Cloud system.}
        \label{tab:temps}
\end{table}

%\begin{figure}

\subsection{Implementation of a cross-environment CFD scientific workflow}
\label{sect:exp-setup}
\begin{figure}[t]
    \centering
    \includegraphics[width = 0.8\textwidth]{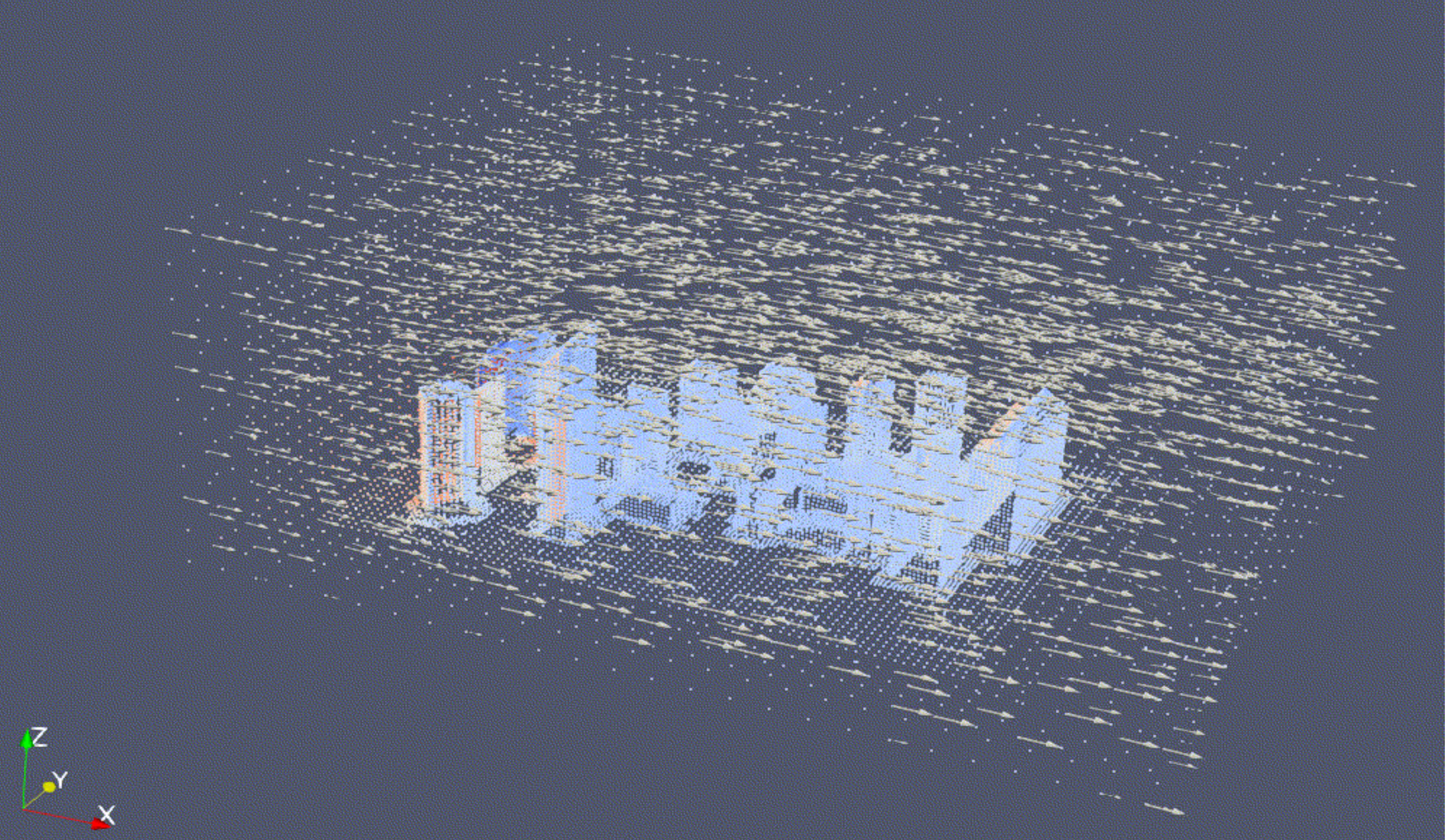}
    \caption{\small Visualization of $WindAroundBuildings$ simulation using ParaView \cite{ahrens2005paraview}.}
    \label{fig:wind-visual}
\end{figure}

Our cross-environment in-situ scientific workflow has two applications: CFD simulation and DMD analysis.
To implement the CFD simulation application, we use the parallel OpenFOAM software \cite{jasak2007openfoam,opencfd2020openfoam}, deployed in IU Karst.
In OpenFOAM, a ``solver'' is the simulation algorithm
%(e.g., using LES \cite{deardorff1970les} or DNS \cite{moin1998dns}),
and a ``case'' describes the physical condition of the simulation problem.
We choose the {\tt simpleFoam} as our solver, which is a steady-state solver for incompressible, turbulent flow, using the SIMPLE (Semi-Implicit Method for Pressure Linked Equations) algorithm.
The simulation problem we choose to solve is the {\tt WindAroundBuildings}, as shown in Figure \ref{fig:wind-visual}.
This case simulates how wind flows behave around a group of buildings in an urban area.
To enable the in-situ workflow execution by using ElasticBroker,
we need to replace the original {\tt runTime().write} function in the {\tt simpleFoam} solver with our {\tt broker\_write} function.
We divide the simulation problem domain into different processes along the $Z$ (height) axis.
The velocity fields of each process region are sent out through the broker, and will be analyzed by the stream processing service deployed in the Jetstream Cloud.

%\begin{figure}
%\begin{subfigure}{0.4\textwidth}

\begin{figure}
    \centering
    \includegraphics[width = 0.8\textwidth, height=80mm]{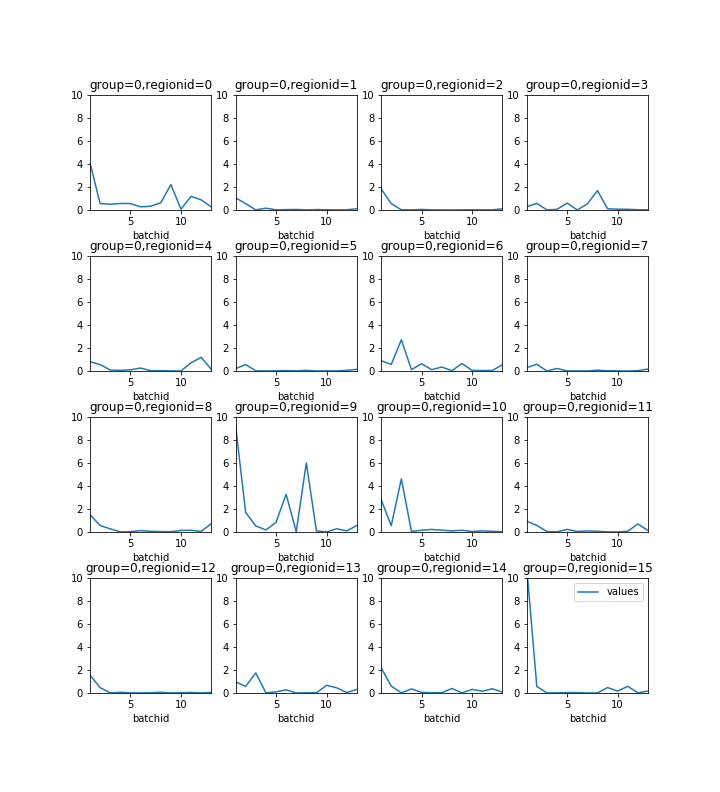}
    \caption{Analysis results of eigenvalues of DMD low-rank operator from each process region's output. Each subplot shows the average sum of square distances from eigenvalues to the unit circle of that region. Values closer to 0 mean fluids in that region are more stable.}
    \label{fig:wind-analysis}
\end{figure}

The analysis application reads data streams from HPC processes through the Cloud endpoints described in Section \ref{sect:hpc-components}.
Figure \ref{fig:wind-analysis} shows the visualization results of DMD analysis on $16$ data streams received by $1$ Cloud endpoints.
Each subplot corresponds to the fluid data sent from one simulation MPI process, and shows how the fluid dynamics change over time for this process region.
This figure can inform users how stable the fluids in each process region is, while the simulation is running.

%\begin{subfigure}{0.6\textwidth}
%\end{subfigure}
%\end{figure}

\subsection{End-to-end workflow time}
One concern of using in-situ processing is that it can slow down simulation applications,
increasing the overall end-to-end time of the workflow.
Traditionally, simulation applications write simulation output to parallel file systems.
The stored files can be used for future post-analysis.
Such file-based I/O is usually expensive, and can also potentially slow down the primary simulation applications.
To investigate how the simulation application and the combined workflow (with Cloud-based DMD analysis) can be affected by different I/O methods, 
we configure the simpleFoam solver (with $16$ processes) in three different modes:
\begin{enumerate}
    \item File-based: simulation output data is written to parallel Lustre file system using the ``collated'' write provided in OpenFOAM.
    \item ElasticBroker: simulation output data is sent to Cloud endpoints, using the proposed ElasticBroker API.
    \item Simulation-only: The simulation runs with data write disabled.
\end{enumerate}

\begin{figure}
    \centering
    \includegraphics[width = 0.9\textwidth]{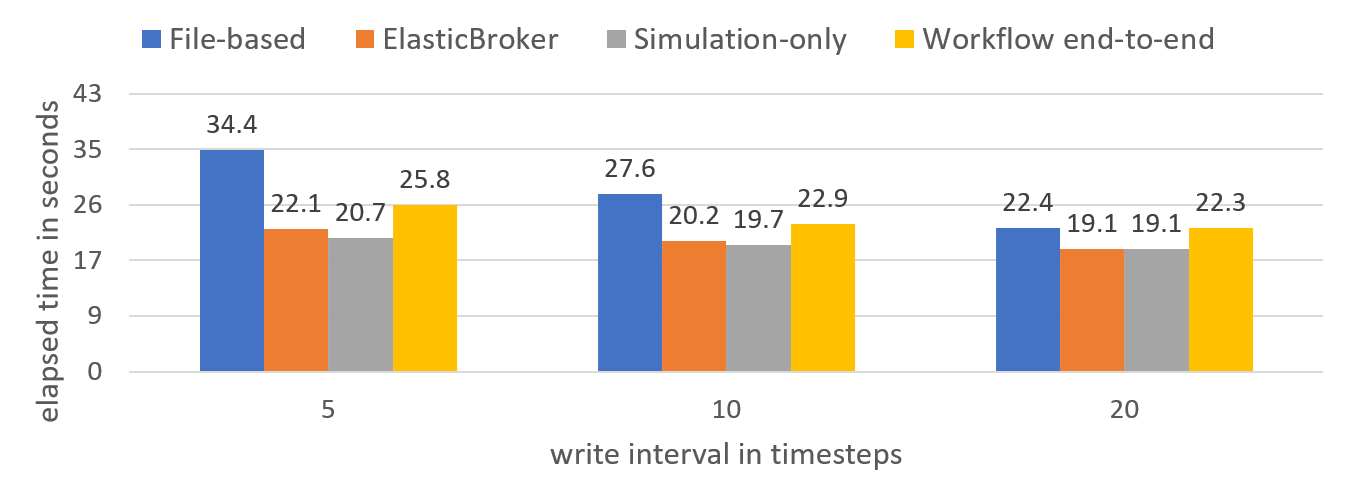}
    \caption{Simulation elapsed time comparison when running \emph{WindAroundBuildings} case with file-based I/O, ElasticBroker and simulation-only.
    The figure shows that while file-based I/O significantly slows down the simulation application, the proposed ElasticBroker method doesn't affect simulation much. The last column shows the end-to-end time of the whole workflow when ElasticBroker is used.}
    \label{fig:exp-end2end}.
\end{figure}

The elapsed time of the simulation application (from simulation starts till simulation finishes) using these different modes are shown in Figure \ref{fig:exp-end2end}.
In the figure, there is also an extra column: the workflow end-to-end time, which starts at the beginning of the CFD simulation and stops
at the end of the DMD analysis in Cloud.
%We run the simpleFoam solver with the \emph{WindAroundBuildings} case in OpenFoam, as introduced in Section \ref{sect:exp-setup}.
We run the simulation application for 2000 time steps (using the configuration of $deltaT=0.2$ and $totalTime=400$ in the OpenFOAM control dictionary file).
To examine how those I/O modes affect simulation applications, we use different write intervals.
For example, with $interval=5$,
the simulation will output simulation results once every 5 timesteps.

From Figure \ref{fig:exp-end2end},
we can see that when the simulation application is configured with long write intervals (e.g. $write\_interval$=20, meaning less-frequent writes), simulation time is similar in different I/O modes.
However, when we increase the write frequency (e.g. $write\_interval=5$),
the file-based method makes the simulation application significantly slower, compared with the simulation-only baseline mode.
In comparison, with ElasticBroker, simulation applications can run with only a minimal slowdown.
This is due to the fact that ElasticBroker asynchronously writes in-process simulation to data streams, from each simulation process, independently.
Compared with the file-based method, no shared file systems are used for output of the bulk simulation, so the simulation can run with much fewer stalls. 

% \todo{how much data is generated?}

In the Cloud side, we configure $16$ Spark-executors deployed in a Kubernetes cluster.
We configure the DMD analysis to be triggered every $3$ seconds for all data streams.
%\todo{can draw the difference of the 3 seconds}
Note that the difference between workflow end-to-end time and the ElasticBroker simulation time in Figure \ref{fig:exp-end2end} is also around $3$ seconds, which means, apart from the configured trigger time, there is no significant lag between simulation and analysis applications.
In conclusion, plugging CFD simulation with ElasticBroker gives us in-time insights of the running simulation,
and it doesn't harm the performance of the simulation much.

\subsection{Throughput}
To better understand the performance behavior of running such workflow in the proposed system,
we conduct a second set of experiments,
in which we illustrate how the system scales when we are using more HPC and Cloud resources.

\begin{figure}
     \centering
     \begin{subfigure}[b]{0.48\textwidth}
         \centering
         \includegraphics[width=\textwidth]{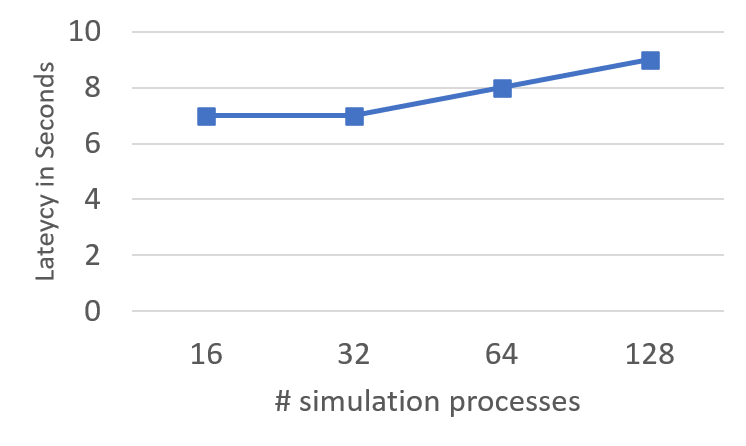}
         \caption{Analysis Latency}
         \label{fig:scale-latency}
     \end{subfigure}
     \hfill
     \begin{subfigure}[b]{0.48\textwidth}
         \centering
         \includegraphics[width=\textwidth]{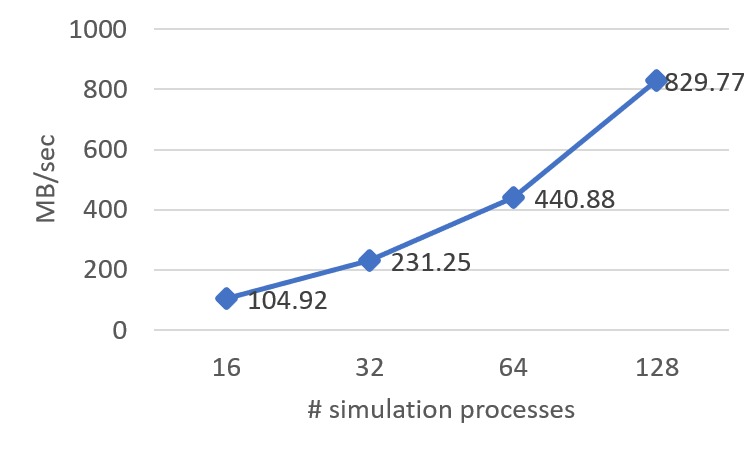}
         \caption{$Throughput$}
         \label{fig:scale-throughput}
     \end{subfigure}
        \caption{Running workflow with a synthetic data generator in HPC, and DMD analysis in Cloud. The number of analysis processes (Spark-executors) is the same as the number of simulation processes.}
        \label{fig:weak-scale}
\end{figure}
Differently from the previous end-to-end time experiments, which use real-world CFD software, we use a synthetic data generator in this part, to produce enough data in order to stress the system.
The synthetic data generator consists of groups of MPI processes in the HPC side.
Data is continuously generated from all processes and streamed to the distributed stream processing service
through multiple Cloud endpoints,
as we have seen in Figure \ref{fig:hpc-components}.
For larger scale simulations, we increase the number of Spark-executors and Cloud endpoints (i.e., Redis server instances) correspondingly.
The ratio among MPI processes, Cloud endpoints, and Spark-executors is set as $16:1:16$.
%as in the simulation time experiment.

We evaluate the latency between analysis and simulations, which is from the time when simulation output data is generated, to the time when the data is analyzed by Cloud services. 
This metric describes the quality of service of our system, which indicates how quickly we can get insights into the running simulations.
From Figure \ref{fig:scale-latency}, we can see the latency stays between $7 \sim 9$ seconds when using $16 \sim 128$ simulation processes.
In Figure \ref{fig:scale-throughput}, which shows the aggregated throughput from all MPI processes, 
we can observe that when we double the number of MPI processes, the aggregated throughput also increases by around two times.
Such scalability benefits from the careful mappings of MPI processes, Cloud endpoints, and Spark-executors.
Specifically, MPI processes in one group always write data to a designated Redis
endpoint, then the data is analyzed by a fixed subset of the Spark-executors.

\section{Related work}
% vs scientific gateways: gateways still uses abstraction of files, can it do analysis "step by step"?
Scientific workflows have been widely used to incorporate multiple decoupled applications running on distributed computational resources.
To manage data dependencies among different applications, and correctly schedule computational tasks, workflow management systems (e.g., Pegasus \cite{deelman2015pegasus}, Kepler \cite{ludascher2006kelpler}) are used.
However, these workflow systems heavily rely on file-based I/O, and
only schedule coarse-grain workflow tasks in a sequential manner (i.e.,  a later task cannot start until all the previous tasks have exited).
In the case of $ElasticBroker$, simulation data is streamed continuously to Cloud services, where data analysis will be conducted while the simulations continue running.

There exist several previous works that deal with different file formats in Big Data and HPC ecosystems.
For instance, LABIOS \cite{kougkas_labios_2019} utilizes the label-based I/O system to bridge HPC and Big Data applications.
NIOBE \cite{kun2019NIOBE} uses I/O forwarding nodes and Burst buffer to stage data and offload the data format conversion operations.
However, these conversion operations still require a shared file system or shared storage system.

Data transport libraries such as ADIOS  \cite{lofstead_flexible_2008}, Decaf \cite{dreher2017decaf}, and Zipper \cite{fu_performance_2018}
do not rely on file-based communications between applications,
but they most often require applications to run in an HPC ecosystem.
Differently, in $ElasticBroker$, data can be sent from HPC applications to endpoints exposed by Cloud services, so that decoupled applications can collaborate while residing in their native environments.

\section{Conclusion and future work}
In this paper, we present our preliminary work that bridges the HPC and Cloud ecosystems and enables cross-system in-situ workflows.
We design $ElasticBroker$, which provides a C/C++ library which MPI applications can be linked to. ElasticBroker can transform simulation data into stream records, and send the stream records to a distributed stream processing service deployed in Cloud.
We also show how the Cloud-based stream processing service is setup, and how it partitions, processes and analyzes the stream data continuously.
We use the parallel OpenFOAM simulation which runs in IU Karst, and DMD analysis which is deployed in XSEDE Jetstream to demonstrate the effectiveness of our framework.
Experimental results show that extending MPI-based simulations with ElasticBroker
enables stream processing services deployed in Cloud to provide in-time analysis of ongoing fluid dynamics.
The experiments also show good throughput and quality of service of ElasticBroker when we increase both simulation and analysis scales.

In our future work, we plan to extend ElasticBroker to support in-situ workflows with more complex directed acyclic graphs (DAG).
More advanced data aggregation functionality can be used in the HPC side so that processes may utilize the bandwidth more efficiently,
Additionally, performance models can be designed to automatically decide how to distribute computation tasks of an in-situ workflow to different environments (e.g., HPC and Cloud), based upon application-specific requirements such as computation time, memory consumption, and migration cost.
%3. extend the broker service, so that we can also provide insights for data generated from edge, e.g. IoT sensors.
%4. Integrate with interfaces like ADIOS2, so more applications can transparently use it.

\section*{Acknowledgement}
This research is supported by the NSF award \#1835817.
This work also used the Extreme Science and Engineering Discovery Environment (XSEDE),
which is supported by NSF grant number ACI-1548562.

%\begin{thebibliography}{6}

%\end{thebibliography}
%\bibliographystyle{plain} %styles/bibtex/unsrtnat}
%\printbibliography
\bibliographystyle{unsrt}
\bibliography{elasticbroker}

\end{document}